\documentstyle[aps,preprint]{revtex}
\begin{document}

\input epsf

\title{On the stability of  
$2 \sqrt 2 \times 2  \sqrt 2$ oxygen ordered superstructures
in YBa$_2$Cu$_3$O$_{6+x}$\\}
\author{A.A. Aligia$^{a,b}$, S. Koval$^{c}$ and R. Migoni$^{a,c}$.}
\address{$^{a}$International Centre for Theoretical Physics, Strada 
Costiera 11, P.O.B. 586, 34100 Trieste, Italy\\ 
$^{b}$Centro At\'{o}mico 
Bariloche and 
Instituto Balseiro, Comisi\'on Nacional de Energ\'{\i}a At\'{o}mica, 
8400 Bariloche, Argentina.\\
$^{c}$Facultad de Ciencias Exactas, Ingenier\'{\i}a y Agrimensura e
Instituto de F\'{\i}sica de Rosario, Universidad Nacional de Rosario, 
Bv. 27 de febrero 210 bis, 2000 Rosario, Argentina}
\maketitle

\begin{abstract}
 We have compared the ground-state energy of several observed or proposed
"$2 \sqrt 2 \times 2 \sqrt 2$ oxygen (O) ordered superstructures " 
(from now on HS), with
those of  ``chain superstructures" (CS) (in which the O atoms of the basal
plane are ordered in chains), for different compositions $x$ in
YBa$_2$Cu$_3$O$_{6+x}$. The model Hamiltonian contains i) the Madelung
energy, ii) a term linear in the difference between Cu and O hole
occupancies which controls charge transfer, and iii) covalency effects
based on known results for $t-J$ models in one and two dimensions. The
optimum distribution of charge is determined minimizing the total energy,
and depends on two parameters which are determined from known results for
$x=1$ and $x=0.5$. We obtain that on the O lean side, only CS are stable,
while for $x=7/8$, a HS with regularly spaced O vacancies added to the $x=1$
structure is more stable than the corresponding CS for the same $x$.
We find that the detailed positions of the atoms in the structure, and 
long-range Coulomb interactions, are crucial for the electronic structure, 
the mechanism of charge transfer, the stability of the different phases, and the
possibility of phase separation. 
\end{abstract}



\section{Introduction.}

The ordering of oxygen (O) atoms in the basal plane of YBa$_2$Cu$_3$O$_{6+x}$,
and its relation with electronic properties, particularly
the superconducting critical temperature $T_c$, has been a subject of
great interest. Overviews are contained in Ref. \cite{rev}.
By the end of the last decade, for $x$ near 0.5 evidence or ordering in
chains was found, while for $x$ near 0 or 1, HS were reported 
\cite{ala,rey,wer,bey}, the charge distribution in superconducting planes 
\cite{wan} and in the Cu ions of the basal planes \cite{tra,war} was determined,
and an explanation of these experiments was presented using a lattice-gas model
based on Coulomb repulsions, and the appropriate extension to this system of the
three-band Hubbard model $H_{3b}$ \cite{ori}. Optical \cite{tra,tol} and nuclear
quadrupole resonance (NQR) \cite{war,lut} experiments give strong evidence that 
two-fold coordinated Cu atoms are in an oxidation state Cu$^+$, while three- and
four-fold coordinated Cu atoms are Cu$^{+2}$. Qualitatively, this
fact is easy to understand in terms of $H_{3b}$
\cite{note}: the energy necessary to add a $d$ hole to an $n$-fold coordinated
Cu$^+$ (surrounded by $n$ O$^{-2}$ ions) is $\epsilon_{Cu} -2nU_{pd}$, where
$U_{pd}$ is the Cu-O hole-hole repulsion. Provided that the hole Fermi energy of
the superconducting CuO$_2$ planes lies between $\epsilon_{Cu} -4U_{pd}$ and 
$\epsilon_{Cu}-6U_{pd}$, only two-fold coordinated Cu atoms remain Cu$^+$, and
the remaining holes go to the CuO$_2$ planes (and to  O atoms of the chains). 
This mechanism not only provides the holes for superconductivity of the
CS, but also reduces their energy \cite{note}.
For $x \sim 0.5$, annealing at room
temperature produces an increase of $T_c$ due to an increase in the amount of
two-fold coordinated Cu ions \cite{vea,cla,kir}.

In spite of this qualitative understanding of the relation between O ordering 
and charge transfer, and the success of lattice-gas models in explaining the 
basic features of the thermodynamics of O ordering at temperatures above room 
temperatures \cite{ori,hil,ter,zub,hau}, a fully self-consistent theory of the 
atomic and electronic structure of  YBa$_2$Cu$_3$O$_{6+x}$ is still lacking, and
several controversial issues remain. Schleger {\it et al.} showed that it is 
necessary to add electronic degrees of freedom to lattice-gas models in order to
explain the observed $\partial x / \partial \mu$, where $\mu$ is the O chemical 
potential \cite{sch}. The strong-coupling approaches to the electronic structure
are able to treat adequately the on-site Cu Coulomb repulsion $U_d$,
and to explain the observed dependences of the hole count in the planes $n_H$ 
(from which $T_c$ can be inferred \cite{vai}), and the amount of Cu$^+$ 
\cite{rev,hau,vai,lat,uim}. However, these agreements are obtained within a 
region of parameters of the model, which is not fully justified, and Coulomb 
repulsions beyond $\sim 2 \AA$ are neglected. On the other hand, {\it ab initio}
calculations, as a consequence of the large value of $U_d$ and the neglect of 
correlations, fail to describe the semiconducting phases (they predict a metal) 
and the neglected correlation energies are much larger than the lattice-gas 
model parameters which determine the structure \cite{rev}.

A controversial issue is the stability of HS. While it is clear that at room
temperatures, the CS disappear in the semiconducting region \cite{lut}, the
experimental and theoretical situation does not allow at present to disclose
unambiguously the nature of the ground state. The HS observed by transmission 
electron microscopy \cite{ala,rey}, were proposed to be metastable \cite{rey}. 
Synchrotron radiation studies for $x \sim 0.2$ provided strong evidence of the 
presence of $ \sim 0.2 \%$ of a parasitic phase BaCu$_3$O$_4$ \cite{yak}, which
is able to explain the x-ray diffraction pattern observed for $x \sim 3/8$, 
ascribed previously to O ordering \cite{sonx}. This pattern is also fully 
compatible with the superstructure of minimum Coulomb repulsion between equally 
charged O atoms, among all those with  unit cell 
$2 \sqrt{2} \times 2 \sqrt{2}$ \cite{sonx} (see Fig. 1). On 
the other hand, neutron-scattering experiments for $x \sim 3/8$ \cite{sonn}, 
which are more suitable to study O ordering due to the comparatively  larger O 
cross section \cite{ero}, are so far only explained in terms of an O ordered 
superstructure of unit cell $2 \sqrt{2} \times 4 \sqrt{2}$ \cite{ero,38}.  On 
the theoretical side, there is no general physical argument from which one can 
discard or confirm HS \cite{rev,ero,sol,v2}, except under too restrictive 
hypothesis \cite{v2}, as we will show in Section III. The model which obtained 
these superstructures \cite{ori,note2} is based on Coulomb repulsions between 
any two O ions, with metallic and dielectric screening and a parameter 
$\Delta E$, which favors CS, to take into account the relation between charge 
transfer and Cu coordination mentioned above \cite{rev,sol}. $\Delta E$ has been
calculated using an extended Hubbard model, but the result is too sensitive to
the parameters of this model, not known with enough precision. For large
but reasonable $\Delta E$, the ground state of the structural model is a CS for 
any O content $x$ \cite{rev,sol}.
    
Because of the very low or vanishing density of carriers, the Madelung energy is
a fundamental ingredient in the physics of  YBa$_2$Cu$_3$O$_{6+x}$
\cite{kon,fei,qui,tor} and other \cite{tor} high-$T_c$ systems. In
YBa$_2$Cu$_3$O$_{6+x}$, O-O repulsions at distances of at least $8 \AA$ are
necessary to explain observed split diffuse diffraction peaks \cite{one}, and 
O-O repulsions at distances $\sim 27 \AA$ are required to stabilize several of 
the observed CS \cite{sol}. Also, the model of Ref. \cite{ori}, based on Coulomb
repulsions, seems to be able to
explain qualitatively the experimental data gathered on  YBa$_2$Cu$_3$O$_{6+x}$
\cite{rev}, including the structure transformation kinetics if supplemented by
long-range elastic energies \cite{sem}. Here we generalize previous works 
considering the Madelung energy, allowing for the possibility of charge
transfer and finding the optimum distribution of charges by minimization of the 
total energy. Effects of covalency are also included. Neglecting the latter, 
our approach is equivalent to the exact solution, in the limit of zero
hopping and infinite on-site Coulomb repulsions, of the appropriate model of
the Hubbard type for the system. The effect of correlations, essential for
Cu-O charge transfer \cite{note} is adequately retained.
The main shortcomings of the approach
are the neglect of core-core repulsive energy (which amounts to $10 \%$ of
the Madelung energy and stabilizes the lattice) and screening effects.

The theoretical treatment is presented in Section II. Section III contains
the results and Section IV is a discussion.

\section{The model}

The ground-state energy of the system per Y atom is described as:
\begin{eqnarray}
E=E_{Mad}+E_{\Delta}+2E_{pl}+E_{ch}.
\end{eqnarray}
\noindent $E_{Mad}$ is the Madelung energy as a function of Cu and
O charges. $E_{\Delta}$ is the
energy required by the charge-transfer process Cu$^+$ + O$^-$
$\rightarrow$ Cu$^{+2}$ + O$^{-2}$ in absence of interatomic repulsions.
$E_{pl}$ describes the kinetic and magnetic energy gain due to covalency
in the superconducting CuO$_2$ planes (assumed equal for both planes of the unit
cell), and $E_{ch}$ is the corresponding
term for the CuO$_{2+x}$ subsystem containing the basal plane.

The Madelung energy can be written as \cite{kov}:
\begin{eqnarray}
E_{Mad}=\frac {e}{2N} \sum_i \beta_i Z_i; ~
\beta_i=e \sum_j \alpha_{ij} Z_j,
\end{eqnarray}
where $e$ is the elementary charge, $N$ the number of Y atoms in the supercell,
$Z_i$ the
charge of the $i^{th}$ atom in the supercell, and $\beta_i$ the
electrostatic potential at the position of this atom.
Because of charge neutrality $\sum_i Z_i=0$, the geometrical
coefficients $\alpha_{ij}$ can be shifted by an arbitrary constant. We have
chosen it in such a way that $\alpha_{ii}=0$ for all $i$. Then for 
$i \neq j$:
\begin{eqnarray}
\alpha_{ij}=\frac{1}{r_{j,{\bf 0}}- r_{i,{\bf 0}}} + 
\sum_{{\bf T \neq 0}} 
(\frac{1}{r_{j,{\bf T}}-r_{i,{\bf 0}}}-\frac{1}{r_{i,{\bf T}}-r_{i,{\bf 0}}}).
\end{eqnarray}
Here ${\bf T}$ labels the translation vectors which map the superlattice onto 
itself and $r_{k,{\bf T}}= {\bf T}+ r_{k,{\bf 0}}$ is the position of the 
$k^{th}$ atom of that supercell obtained from the one lying at the origin by a 
translation ${\bf T}$.
The sum over  ${\bf T}$ is evaluated by the Ewald's method \cite{wei}. For
simplicity we have assumed that the lattice parameters $a=b$ (taking the average
between them), and we have taken the positions of the atoms from three different
O contents:$x=7$ \cite {gui}, $x=0.45$ \cite {cav} and $x=0$ \cite {cav}. We 
also assume that all Y ions are Y$^{+3}$, all Ba ions are Ba$^{+2}$, all Cu ions
of the superconducting planes (Cu(2) in the notation of Refs. \cite {gui,cav}) 
have the same charge, and all O atoms of these planes (denoted O(2), O(3) 
\cite{gui,cav}) related by translations of the primitive $1 \times 1$ cell have 
the same charge. This simplifies considerably the problem, allowing us to 
express part of the sums in $i$ and $j$ of Eqs. (2) and (3), in terms of the 
coefficients $\alpha_{ij}^0$ of the primitive unit cell, reducing appreciably
the number of $\alpha$'s which should be calculated.
 Neglecting an unimportant constant we can write:
\begin{eqnarray}   
E_{\Delta}=3 \Delta (Z_{Cu}-2),
\end{eqnarray}
where $Z_{Cu}$ is the average charge of all Cu atoms. In principle, $\Delta$ is 
the difference between the ionization potential of Cu$^+$ and the (negative) 
electron affinity of O$^-$. However, it should also contain information of 
steric effects (short-range repulsions), the energy gain of a {\em local}
Zhang-Rice singlet \cite{ero} (not included in $E_{pl}+E_{ch}$) and the kinetic
energy of the Cu hole. We keep $\Delta$ as a parameter, assumed independent of 
$x$ and determined in such a way that the charge distribution in  
YBa$_2$Cu$_3$O$_{6.5}$ agrees with experiment.

The number of added holes in one of the two superconducting planes per unit cell
is $h=2+Z_{Cu2}+Z_{O2}+Z_{O3}$, where $Z_{Cu2},~Z_{O2},~Z_{O3}$ are the average 
charges of the Cu(2), O(2) and O(3) atoms of that plane. For the kinetic energy 
as a function of $h$, we take the form established from a high-temperature 
expansion of the $t-J$ model, with $t$=0.4  eV, $J$=0.1  eV \cite{feh}, slightly
generalized to give the correct magnetic energy for $h=0$:
\begin{eqnarray}   
E_{pl}=-[\epsilon h + (1-h)(h+0.09192) eV].
\end{eqnarray}
$\epsilon$
represents the difference between the energy gain in forming a localized 
Zhang-Rice singlet in the planes with respect to forming it in perfect CuO$_3$ 
linear systems (involving the Cu(1), O(4) atoms of the basal plane, and two O(1)
apical atoms \cite{gui,cav}). It can also contain information of different 
steric effects in chains and planes. We keep $\epsilon$ as a parameter determined
from the experimental charge distribution for $x=1$. Actually, a realistic 
one-band model for the cuprates contains also hoppings beyond nearest neighbors 
and three-site terms which determine the shape of the Fermi surface and are 
critical for the superconductivity \cite{fe,ba}, but we expect that these terms 
do not affect $E_{pl}$ very much.  

$E_{ch}$ is obtained by fitting exact results for the one-dimensional $t-J$ 
model. The values $t=0.85$ eV, $J=0.2$ eV were determined from a
low-energy reduction procedure which leads to excellent results for the optical
conductivity of the chains \cite{gag}. The result is \cite{gag,v2}:
\begin{eqnarray}
E_{ch}=[(J-2t)\sin (\pi (1-h_c))-J(0.69+0.41 h_c)(1-H_c)]y_1 - J y_2,
\end{eqnarray}
where $y_1$ is the concentration of O(4) atoms belonging to perfect chains, $y_2$
is the concentration of isolated O(4)$^{-2}$ ions between two Cu(1)$^{+2}$ ions,
and $h_c$ is the number of holes per Cu,  added to (CuO$_3)^{-4}$ perfect chains.
This expression does not take into account doped extremely short chains or the 
correction corresponding to chains of intermediate length.
 
 The total energy $E$ is minimized with respect to the charges in the subsystem
CuO$_{2+x}$ (containing Cu(1), O(1) and O(4) atoms), and the average Cu(2), O(2),
O(3) charges on both superconducting planes. Cu charges are allowed to vary 
between 1 and 2, and O charges between -2 and -1. This corresponds to the limit 
of very large on-site Coulomb repulsions. 

\section{Results.}

First, we have applied our approach to stoichiometric YBa$_2$Cu$_3$O$_7$, a 
metallic state with a reasonably well-known, non trivial charge distribution. 
This is difficult to obtain from the Madelung energy, because this concept was 
developed for insulators. In fact, if we neglect  $E_{pl} and E_{ch}$, we obtain
that all Cu ions are Cu$^{+2}$, and all O ions are O$^{-2}$ except the chain O(4)
atoms, the oxidation state of which is -1. In other words all holes go to the
one-dimensional Cu(1)-O(4) chains which are not expected to conduct (due to 
defects or Peierls distortions) and the system is insulating. This is not bad as
a first approximation. It is the best description of the observed charge 
distribution in terms of integer charges. Including all terms in the energy, 
with $\Delta < 46$ eV and a reasonable $\epsilon=2$ eV, we obtain that $60\%$ of
the holes enter CuO chains and  $20\%$ are in each of the superconducting 
CuO$_2$ planes, in agreement with optical conductivity measurements \cite{zsch},
and estimations based on bond valence sums and other experiments \cite{tal}. The
resulting charges, Madelung potentials $\beta_i$ and different contributions to 
the energy are included in Table I.

After checking that in general this gives lower energy, and to simplify the
algorithm, we have constrained the minimization procedure distributing the holes
on both superconducting planes in equal amounts between the O(2) atoms of one 
plane and the O(3) of the other, and have kept -2 the charge of the apical O(1) 
ions. The results depend significantly, but not dramatically on $\epsilon$: for 
$\epsilon$=0, the amount of holes in each plane is reduced to $14\%$. Instead, 
the positions of the Ba and apical O atoms are crucial. 
If the atomic positions are taken as those for $x=0$, for which the O(1)
atoms are nearer and the Ba atoms more distant from the basal plane, we obtain a
hole doping $h$ of only 0.03 in each CuO$_2$ plane, while for the positions
corresponding to $x=0.45$, the resulting doping is 0.14.     

To establish bounds on $\Delta$, we have calculated next the energy and charge
distribution of both superstructures shown in Fig. 1 for $x=1/2$. The 
HS is the one which minimizes the Coulomb
energy when all atoms related by symmetry operations of the tetragonal primitive
unit cell have the same charge \cite{gro,c,note2}. However, the ground state is 
the CS, the two-fold (four-fold) coordinated Cu ions are mainly Cu$^+$ 
(Cu$^{+2}$) \cite{tra,tol,war,lut}, and about 0.1 holes per Cu go to the 
superconducting CuO$_2$ planes \cite{wan,tal}, in agreement with theory 
\cite{rev,hau}. To satisfy this charge distribution, our model has to satisfy 
several constraints. One of them is 
$\Delta > \Delta'_{min}=e(\beta_{Op}-\beta_{Cu2})- \partial E_{pl} / \partial h$, 
where $\beta_{Op}$ is the lowest $\beta_i$ of the occupied O atoms of the
planes, and $\beta_{Cu2}$ is the potential at the two-fold coordinated chain 
Cu(1) atoms. If this constraint is not satisfied, the holes of the planes go to 
the two-fold coordinated Cu atoms and the system would be insulating. In Table 
II we give the resulting charges, $\beta_i$ and energies of the CS for 
$\Delta=31$ eV, slightly above  $\Delta'_{min}$. For larger values of  $\Delta$ 
(not too large to avoid that four-fold coordinated Cu$^{+2}$ becomes Cu$^+$), 
the only change is that $E$ and $E_\Delta$ decrease proportionally to 
$\Delta/2$. The resulting amount of holes in each CuO$_2$ plane (0.093 per Cu) 
is in very good agreement with experiment. The $\beta_i$ at inequivalent O(1) 
atoms are surprisingly similar and a little bit smaller than the $\beta_i$ of 
the O atoms of the CuO$_2$ planes. Holes prefer the latter because of the 
positive value of $\epsilon$. However a more realistic description should allow 
that a small amount of holes enter apical O(1) atoms, particularly for small  
$\epsilon$.

The experimental evidence indicates that three-fold coordinated Cu(1) atoms are
Cu$^{+2}$, in particular optical experiments on quenched samples \cite{tra,tol},
and NQR experiments in which Y is replaced by larger ions \cite{lut}. This is 
also suggested by the theoretical studies \cite{rev,hau,v2} and the argument on 
the interplay between charge transfer and structure \cite{note} presented in 
Section I. An upper bound on $\Delta$ can be obtained requiring that in the HS 
for $x=1/2$, all Cu ions remain Cu$^{+2}$ and no holes are transfered to the 
CuO$_2$ planes, so that the system remains semiconducting. This implies   
$\Delta < \Delta_{Max}=e(\beta_{Op}-\beta_{Cu(1)}) - \partial E_{pl} / \partial h|_{h=0}$. 
We obtain $\Delta_{Max}=38.29$ eV. It is interesting to note that in order for 
the present work to be consistent with the estimations of Ohta {\it et al.} for 
the charge transfer gap in several high-$T_c$ superconductors \cite{oht} one 
should take $\Delta \sim 3.5 \times 10.9$ eV = 38.15 eV. The results for any 
$\Delta < \Delta_{Max}$ are presented in Table III. Comparison with the energy 
of the CS (Table II), establishes a better lower bound for $\Delta$: in order 
for the CS to be the ground state, $\Delta > \Delta_{min}=31.34$ eV.

In the following, we assume that $\epsilon=2$ eV  and 
$ \Delta_{min} < \Delta < \Delta_{Max}$ for all $x$. Using these criteria we 
derive conclusions regarding the stability of HS in comparison with CS. Let us 
begin with $x=1/8$, calculated with the atomic positions for $x=0$ \cite{cav}. 
For CS, all four-fold coordinated Cu(1) ions remain Cu$^{+2}$, the two-fold 
coordinated ones are Cu$^+$, and most of the holes apported by the neutral O 
atoms entering the $x=0$ structure to form that of $x=1/8$, remain in their 
neighborhood: one hole is transfered to a nearest neighbor Cu$^+$ and only 0.09 
additional holes per supercell are distributed in the planes. In other words the
charge of the chain O(4) atoms is -1.090 and the doping of each superconducting
plane is $h=0.0056$. The potential at the four-fold coordinated Cu(1) atom in 
the supercell is -24.55 V, while those at two-fold coordinated Cu(1) atoms vary 
between -13.24 V to -12.36 V, with increasing distance to the Cu(1)-O(4) chain. 
The difference of more than 11 V is not taken into account in Hubbard-type 
models which do not include a large nearest-neighbor Cu-O repulsion $U_{pd}$ in 
an appropriate way \cite{rev,ori,v2}. $\beta_i$ at the chain O(4) atoms is 
13.12 V. The $\beta_i$ at the CuO$_2$ planes and apical O atoms have similar 
values as those reported in Tables I and II.

In the HS for $x=1/8$, all O ions are O$^{-2}$ and all three-fold coordinated 
Cu(1) ions are  Cu$^{+2}$. $\beta_i$ at these atoms is -23.82 V, while at the 
remaining, two-fold coordinated Cu(1)$^+$ ions, it is $\sim -12$V. At the O(4)
ions $\beta_i= -18.30$ V and other $\beta_i$ are similar as those in Table III. 
For $\Delta = \Delta_{min}=31.34$ eV, the total energy of the CS is -296.33 eV, 
slightly less than that of the HS (-296.07 eV). Although the Madelung energy of 
the latter is less (-272.57 eV in comparison with -268.63 eV of the CS), the
CS has lower energy because it ionizes half of the Cu(1)$^+$ ions of the $x=0$
structure, and thus, pays less  $\Delta$. For larger values of  $\Delta$, the
difference increases linearly with  $\Delta /8$. Thus, the CS is the ground 
state for $x=1/8$.

Under some general conditions, using a multiband Hubbard model including 
$U_{pd}$, one of us has shown that HS-type superstructures have less energy in 
the semiconducting phase if no holes enter apical O(1) atoms \cite{v2}. What is 
the reason of the discrepancy with the present result? On one hand, the effect 
of repulsions beyond nearest neighbors is important. For example, in usual 
Hubbard-type models, the energy necessary to add a hole in a chain O(4) atom 
with both nearest neighbors being  Cu$^{+2}$ is $\epsilon_p + 2U_{pd}$
independently of the rest of the electronic and atomic structure. However, as
explained above, this energy is 13.12 eV for the CS, and 5.28 eV larger for the
HS. On the other hand, to obtain the present charge distribution of the CS with 
the model of Ref. \cite{v2}, $\epsilon_p < \epsilon_d + 2U_{pd}$ is required, 
contrary to what is expected in CuO$_2$ planes \cite{hyb,gra}, and one of the 
hypothesis of Ref. \cite{v2}. As stated clearly in Ref. \cite{v2}, that 
calculation was aimed to discuss the effects of covalency neglecting repulsions 
beyond nearest-neighbor Cu-O ones and assuming that these were small. However, 
we find that longer range repulsions are essential.   

Next, we analyze the superstructures corresponding to $x=3/8$ with atomic 
positions taken from data for $x=0.45$. The general trends of the charge 
distribution and potentials are similar to those of $x=1/8$. Two-fold 
coordinated Cu(1) ions remain Cu$^+$, while higher coordinated Cu ions have an 
oxidation state Cu$^{+2}$. For CS, the charge of chain O(4) atoms is -1.37 and 
the doping per Cu of the superconducting planes is $h=0.07$. The potential at 
the sites of two O(4) atoms of the unit cell is 15.01 V and 16.12 V at the 
remaining O(4) sites. The $\beta_i$ at four-fold coordinated Cu(1) sites are 
near -25 eV, and those at two-fold coordinated Cu(1) sites vary between -11.45 V
and -13.17 V. For the HS, all O ions are O$^{-2}$, the $\beta_i$ at O(1) atoms 
is near 20.8 V, and those at four-fold (two-fold) coordinated Cu(1) atoms are 
near -22 V (-10.35 V). We should note that keeping this chargedistribution, 
there are at least two superstructures with less energy \cite{38}, one of them 
which provides the best fit of the neutron scattering data \cite{38,sonn}. The 
difference between the Madelung energy of the superstructure of Fig. 1 and the 
lowest lying of the above mentioned superstructures is 0.17 eV \cite{38}. 
Including this correction, the energy of the lowest lying superstructure of unit
cell multiple of $2 \sqrt{2} \times 2 \sqrt{2}$ for 
$\Delta = \Delta_{min}=31.34$ eV becomes -294.62 eV, only slightly smaller than 
the energy for the CS (-294.56 eV). Since increasing $\Delta$ favors the CS by a 
term proportional to $3 \Delta /8$, there is a crossing already at 
$\Delta_c=31.50$ eV and for $\Delta_c < \Delta < \Delta_{Max}$ = 38.29 eV, the 
CS has lower energy. If the $x=3/8$ superstructures are analyzed with atomic 
positions corresponding to $x=0$ instead of $x=0.45$, the same trends are 
observed. $\Delta_c$ increases slightly to 31.85 eV. The doping of the planes 
for CS is reduced to $h=0.02$. The energy decreases in $\sim 0.6$ eV for both 
structures. 
This difference becomes significant when one considers the possibility of
phase separation: the structures with $x=3/8$ calculated with the positions for
$x=0$ ($x=0.45$) \cite{cav} are stable (unstable) against phase separation into
phases with $x=1/8$ (calculated with the positions for $x=0$) and $x=1/2$
(calculated with the positions for $x=0.45$). Since we do not know exactly all
atomic positions at the compositions of interest, we cannot draw definite
conclusions regarding phase separation. However, as pointed out earlier 
\cite{sepa}, the relaxation of the lattice is very important and favors phase 
separation.

The last comparison between the two types of superstructures we make is for
$x=7/8$. Since we do not include covalent corrections for Cu(1)-O(4) chains of 
intermediate length present in the HS (see Fig. 1), we drop $E_{pl}+E_{ch}$ in 
this comparison. According to Table I, the magnitude of the neglected terms is 
$\sim 3$ eV. As in previous cases, only two-fold coordinated Cu ions are +1 and 
the rest are  Cu$^{+2}$. As a consequence of the neglect of covalency, all O
atoms of CuO$_2$ planes are O$^{-2}$, and all chain O(4) atoms are O$^-$, except
one of the four O(4) atoms of the $2 \sqrt{2} \times 2 \sqrt{2}$ unit cell, 
nearest to the additional O(4) vacancy, which is  O$^{-2}$ (we have chosen it 
inside the short chain). For the CS, $\beta_i$ at two-fold coordinated Cu(1)$^+$
is -9.14 V and that at four-fold coordinated Cu(1)$^{+2}$ lies near -19.5 V. The
potential at all apical O(1) (chain O(4)) ions lie near 22 V (16.7 V), with 
little variation with distance to the Cu(1)-O vacancy chains. For HS, $\beta_i$ 
at four-fold coordinated  Cu(1)$^{+2}$ ions is -23.10 V, except at the one 
nearest-neighbor to the  O(4)$^{-2}$, which amounts to -27.48 V. At the 
three-fold coordinated  Cu(1)$^{+2}$ ion nearest-neighbor to the  O(4)$^{-2}$ 
ion, $\beta_i$ is also -23.10 V, and at the other three-fold coordinated  
Cu(1)$^{+2}$ ion of the unit cell, $\beta_i= -18.73$ V. 
At the O(4)$^{-2}$ ions is $\beta_i=17.57$ V. The potential at the other
O(4) sites (occupied by O$^{-1}$ ions) vary between 12.42 V and 14.57 V.

The energy of the HS for $x=7/8$ is $E=E_{Mad}=-292.44$ eV. This is less than 
the energy  $E=E_{Mad}+E_{\Delta}$ for the CS, even at the largest possible 
$\Delta = \Delta_{Max}= 38.29$ eV, for which $E= -291.40$ eV. Including  
$E_{pl}+E_{ch}$ this energy decreases to -292.92 eV, but in principle one 
expects a similar decrease for the HS. Also for $\Delta = \Delta_{min}=31.34$ 
eV, even including $E_{pl}+E_{ch}$, the energy of the CS is -291.85 eV, larger 
than that of the HS. We conclude that our model supports the latter as the
ground-state superstructure.

\section{Discussion.}

 We have studied the interplay between the electronic and atomic structures of 
YBa$_2$Cu$_3$O$_{6+x}$ by a novel approach based on the Madelung energy {\em and}
the cost of the charge transfer process Cu$^+$ + O$^-$ $\rightarrow$ Cu$^{+2}$ +
O$^{-2}$. The effect of covalency is included as a correction. This  approach is
motivated by the fact that while first-principles calculations fail to describe the
semiconducting systems, and the correlation energy they neglect is near
0.6  eV and depends on the oxygen ordering \cite{rev}, the strong-coupling models
used so far \cite{rev,hau,vai,lat,uim} neglect long-range repulsions and
depend on
parameters which are not well known. A particular
difficulty of these Hubbard-type models when applied to defects or systems with low
symmetry (as the $1 \times 8$ or $2 \sqrt 2 \times 2 \sqrt 2$ unit cells) is that
they require a large number of uncertain parameters (on-site energies at sites
non-equivalent by symmetry for example) to describe the problem accurately enough.
It is also difficult to include long-range repulsions in exact Lanczos
diagonalizations of Hubbard like models. This was done by Riera and Dagotto for a
generalized three-band Hubbard model in one and two dimensions \cite{ri}. However
the two-dimensional results are incorrect due to subtleties in the use of boundary
conditions, and in general different cluster sizes and shapes should be used
\cite{ric}. 

We addressed the issue of the stability of "chain structures"
(CS) of unit cell  $1 \times n$ ($n$ integer) in comparison with 
$2 \sqrt 2 \times 2 \sqrt 2$ type of superstructures (HS). 
As mentioned in Section I, for
$x < 0.4$ and room temperatures, the experimental evidence is
against CS. We mention here also photoconductivity
experiments \cite{nie,kud,osq,has}: illuminating semiconducting (non CS) films,
the resistivity decreases strongly as a consequence of pumping holes to the
superconducting CuO$_2$ planes, and ordering in (presumably short) chains takes
place \cite{osq} since these structures are energetically favored under the
constraint of a sizeable hole occupancy in the planes (this can be inferred from the
information on the different potentials given in the previous section, or the
arguments given in Refs. \cite{rev,com}). When illumination ceases, the
resistivity returns to the original high values in times characteristic of oxygen
diffusion (see Fig. 2 of Ref. \cite{nie}), showing that the true equilibrium state
is not a CS. Nevertheless, it is still possible that at lower temperatures a phase
transition takes place (difficult to detect because of the sluggish oxygen kinetics
at low temperatures), and the ground state is a CS. The present results support this
statement. It is reasonable to expect that HS are favored by entropy at moderate 
temperatures: for CS,
the cost in energy for a displacement of an O atom to their nearest available
positions, breaking the chains, is high, while this is not the case for HS 
\cite{38}. In fact this entropy term is
essential to explain the neutron-diffraction results for $x \sim 3/8$  \cite{sonn} in
terms of a $2 \sqrt 2 \times 4 \sqrt 2$ superstructure \cite{38}.

The structural model of Aligia, Garc\'es and Bonadeo \cite{rev,sol,c} is based on
Coulomb repulsions between any two basal-plane O(4) ions, screened by free carriers
and dielectric polarization. Except for high stabilization energy of the chains
$\Delta E$, CS are unstable within this model for $x \sim 1/8$ because of the large
cost in O-O Coulomb energy required to arrange the O(4) atoms in Cu(1)-O(4) chains.
The present results show that the neglect of electronic screening
in the semiconducting phase is incorrect: for $x=1/8$ and CS, the resulting O(4)
charge ($\sim -1$), is screened by their nearest-neighbor Cu(1)$^{+2}$ ions,
and the cost in Coulomb energy to build the CS is not so high. Instead, we obtain
that the cost of putting a line of O {\em vacancies} in the $x=1$ structure is not
so efficiently screened and thus, for $x \sim 1$, CS are not favorable.
Other results which should be revised concern the hole count in the planes and the
60K plateau in the superconducting critical temperature
\cite{rev,ori,hau,vai,lat,uim}. The present results suggest that the role of apical
O(1) atoms is not so important as previously assumed \cite{rev,ori,vai}, and that
the positions of the atoms and relaxation of the lattice are crucial in the charge
balance, and also in a possible separation in phases with different oxygen contents
\cite{sepa}.

Dielectric screening and that of free carriers,
neglected in the present approach, are likely to play an essential role in the
problem. One of the attempts to include Madelung potentials in electronic
calculations is that of Ohta, Tohyama and Maekawa \cite{oht}. The authors screened
those potentials by the optical dielectric constant (assumed 3.5 for all systems) to
obtain different parameters of a multiband Hubbard model for the superconducting
Cu-based perovskites. Using these parameters, they have obtained charge-transfer
gaps, exchange constants and other information in good agreement with experiment.
One would be tempted to extend trivially this formalism to calculate the total
energy of YBa$_2$Cu$_3$O$_{6+x}$, dividing all Madelung contributions by 3.5. This
is clearly incorrect, since dielectric and metallic screening should {\em lower} the
total energy of the system (increasing its absolute value) with respect to the
unscreened case, in spite of the fact that the magnitude of the effective
interaction between two defect charges at a distance large in comparison with 
the interatomic
distance, is reduced by dielectric or metallic screening \cite{rev,sol}. 
A simple electrostatic calculation involving two charges surrounded
by a small void sphere (to avoid divergences) in a dielectric medium, shows that
the interaction of the defect charges with the immediate neighborhood causes the
largest reduction of the total energy. Local distortions around added or vacant
O(4) atoms in YBa$_2$Cu$_3$O$_{6+x}$ were calculated by Baetzold \cite{bae} and are
significant. In the semiconducting phase, for which the effect of free carriers can
be neglected, a formalism which takes into account Madelung energies, atomic
potentials and atomic polarizations exists \cite{sep}, and might be applied to
YBa$_2$Cu$_3$O$_{6+x}$ for $x < 0.4$, as an extension and improvement of the present
approach.

\section*{Acknowledgments.}
One of us (S.K.) is supported by the Consejo
Nacional de Investigaciones Cient\'{\i}ficas y T\'ecnicas (CONICET),
Argentina. A.A.A. and R.M. are partially supported by CONICET.


\begin{table}[h]
\begin{tabular}{l|ll}

 & $Z_i$ & $\beta_i$ (V) \\ \hline
Y&3&-29.39\\
Ba&2&-18.41\\ \hline
Cu(2)&2&-26.22\\
O(2)&-1.80&20.04\\
O(3)&-2&21.73\\
O(2)&-2&21.76\\
O(3)&-1.80&20.01\\ \hline
Cu(1)&2&-24.35\\
O(1)&-2&20.11\\
O(4)&-1.39&15.87\\ \hline
$E$&-291.32& eV\\
$E_{Mad}$&-288.59& eV\\
$E_{\Delta}$&0& eV\\
$E_{pl}$&-1.24& eV\\
$E_{ch}$&-1.49& eV\\

\end{tabular}
\end{table}

{\bf Table I}: Charges ($Z_i$), potentials ($\beta_i$) at the different
atomic sites, and different contributions to the total energy of
YBa$_2$Cu$_3$O$_{6+x}$ for $x=1$, $\epsilon=2$ eV and $\Delta < 46.05$ eV.

\newpage

\begin{table}[h]
\begin{tabular}{l|ll}

 & $Z_i$ & $\beta_i$ (V) \\ \hline
Cu(2)&2&-26.73\\
O(2)&-1.91&20.34\\
O(3)&-2&21.15\\
O(2)&-2&21.16\\
O(3)&-1.91&20.32\\ \hline
Cu$_2$(1)&1&-12.23\\
Cu$_4$(1)&2&-24.94\\
O$_2$(1)&-2&20.18\\
O$_4$(1)&-2&20.14\\
O(4)&-1.37&15.59\\ \hline
$E$&-293.86& eV\\
$E_{Mad}$&-276.93& eV\\
$E_{\Delta}$&-15.5& eV\\
$E_{pl}$&-0.71& eV\\
$E_{ch}$&-0.73& eV\\

\end{tabular}
\end{table}

{\bf Table II}: Same as Table I for the CS of $x=1/2$. The subscript of
Cu(1) ions refer to its coordination, and that of apical O(1) ions is the
coordination of its nearest neighbor Cu(1). Parameters are
$\epsilon=2$ eV and $\Delta = 31$ eV.   

\newpage

\begin{table}[h]
\begin{tabular}{l|ll}

 & $Z_i$ & $\beta_i$ (V) \\ \hline
Cu(2)&2&-28.05\\
O(2)&-2&19.79\\
O(3)&-2&19.81\\ \hline
Cu(1)&2&-21.40\\
O(1)&-2&21.78\\
O(4)&-2&21.25\\ \hline
$E$&-294.03& eV\\
$E_{Mad}$&-293.93& eV\\
$E_{\Delta}$&0& eV\\
$E_{pl}$&0& eV\\
$E_{ch}$&-0.1& eV\\

\end{tabular}
\end{table}

{\bf Table III}: Same as Table I for the HS (``herringbone" \cite{yak}) for 
$x=1/2$, and $\Delta < 40.29 -\epsilon$.

\newpage

\begin{figure}[ht]
\begin{center}
\hskip 0.cm
\epsfysize=18cm
\epsfxsize=12cm
\leavevmode
\epsffile{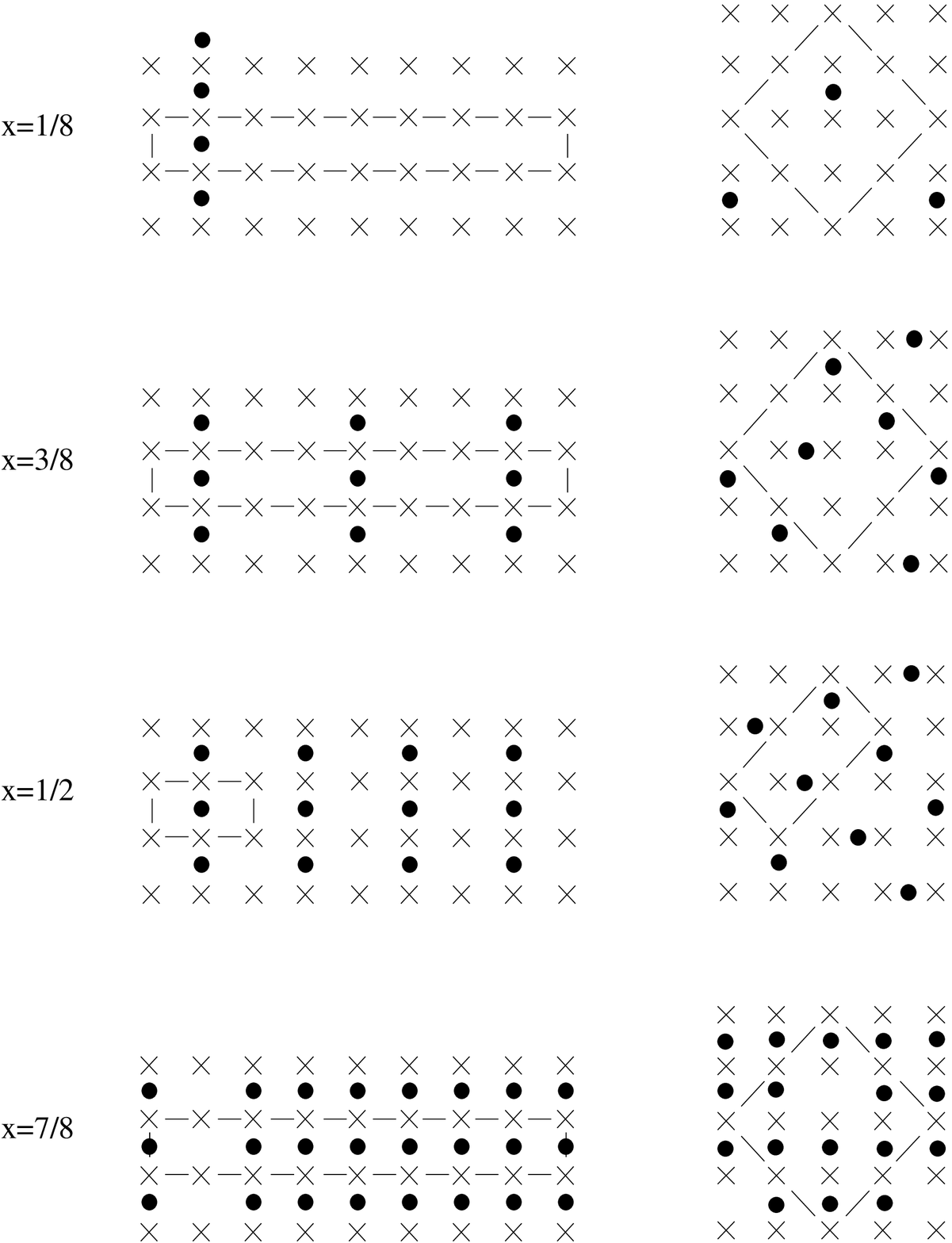}
\vskip 0.cm
\end{center}
\end{figure}

{\bf Fig. 1.} Oxygen ordered superstructures of the basal plane considered
in this work, for different oxygen contents $x$. Left: CS, of unit cell 
$1 \times n$. Right: HS, of unit cell $2 \sqrt{2} \times 2 \sqrt{2}$. Crosses 
denote Cu(1) atoms and solid circles represent O(4) atoms in the notation of 
Refs. \cite{gui,cav}.


\begin{references}
\bibitem{note} This argument for the interplay between Cu coordination, charge
transfer to the superconducting planes and stabilization of ordering in chains, was
introduced in Ref. \cite{gro} (se Fig. 1 of  Ref. \cite{gro}) and is explained in
more detail in Ref. \cite{rev}.
\end{references}
\end{document}